\documentclass[twocolumn,showpacs,amssymb,aps]{revtex4}
\usepackage{amsmath}
\usepackage{amsfonts}
\usepackage{amssymb}
\usepackage{graphicx}
\usepackage{epsfig}
\usepackage{dcolumn}
\usepackage{bm}

\def\lessim{\lower.5ex\hbox{$\; \buildrel < \over \sim \;$}}
\begin{document} \hbadness=10000
\topmargin -0.8cm
\preprint{}
\title{Resonance Production in Heavy Ion Collisions:\\Suppression of $\Lambda(1520)$ and Enhancement of $\Sigma(1385)$}
\author{Inga Kuznetsova$^1$, and Johann Rafelski$^{1,2}$}
\affiliation{$^1$Department of Physics, University of Arizona, Tucson, Arizona, 85721, USA}
\affiliation{$^2$Department f\"ur Physik der Ludwig-Maximilians-Universit\"at M\"unchen and
Maier-Leibnitz-Laboratorium, Am Coulombwall 1, 85748 Garching, Germany}
\begin{abstract}
We investigate the yield of $\Lambda(1520)$ resonance in heavy ion
collisions within the framework of a kinetic master equation without
the assumption of chemical equilibrium. We show that reactions such
as $\Lambda(1520)+\pi \leftrightarrow \Sigma^*$ can favor $\Sigma^*$
production, thereby decreasing the $\Lambda(1520)$ yield.  Within
the same approach we thus find
 a yield  enhancement  for  $\Sigma(1385)$ and a yield suppression for $\Lambda(1520)$.
\end{abstract}

\pacs{24.10.Pa, 25.75.-q, 25.75.Nq, 12.38.Mh}
\date{\today}
 \maketitle

\section{Introduction}
Hadron resonances are observed in a surprisingly large yield when a
quark-gluon plasma  (QGP) fireball breaks up into
hadrons~\cite{Markert:2002xi,Adams:2006yu,Salur:2006jq,Markert:2007qg,Witt:2007xa,Abelev:2008yz}.
This is unexpected, since  the invariant mass signature formed from
decay products could be erased by  rescattering of the strongly
interacting decay products~\cite{Rafelski:2001hp}. Thus  a direct
detection of resonances implies an exceedingly  short period of
hadron scattering, and/or  a final state repopulated by hadronic
interactions~\cite{Bleicher:2002dm}. As a result, the final
resonance yield can be considerably different from statistical
hadron gas (SHG) benchmark expectation. It has already been reported
that the short lived  (compared to characteristic hadron phase
evolution times) resonances  are in general
enhanced~\cite{Kuznetsova:2008zr} compared to SHG benchmark yield.

The new result, we obtain here, is that the long lived  resonances, such as
$\Lambda(1520)$, can be considerably suppressed in their yield. This effect is
amplified for the case when the initial  hadron fugacities, and thus
particle yields,  are  above chemical equilibrium. This situation is
expected for a hadronizing QGP phase. The low
$\Lambda(1520)$ yield  has been  reported
both in RHIC and SPS experiments~\cite{Markert:2002xi,Adams:2006yu}.

In a global QGP breakup (hadron chemical freeze-out) particles and
resonances are formed. Many resonances have a relatively large decay
rates. This implies a  large scattering formation rate. Thus
resonances are  exceptionally strongly interacting particles and
continue to evolve in what we call kinetic phase, even after all
other particles freeze-out. This continued reaction phase is
specific to the resonances and can last considerably beyond the last
non-resonant (elastic) scattering.

\begin{figure*}
\centering
\includegraphics[width=17 cm, height=12 cm]{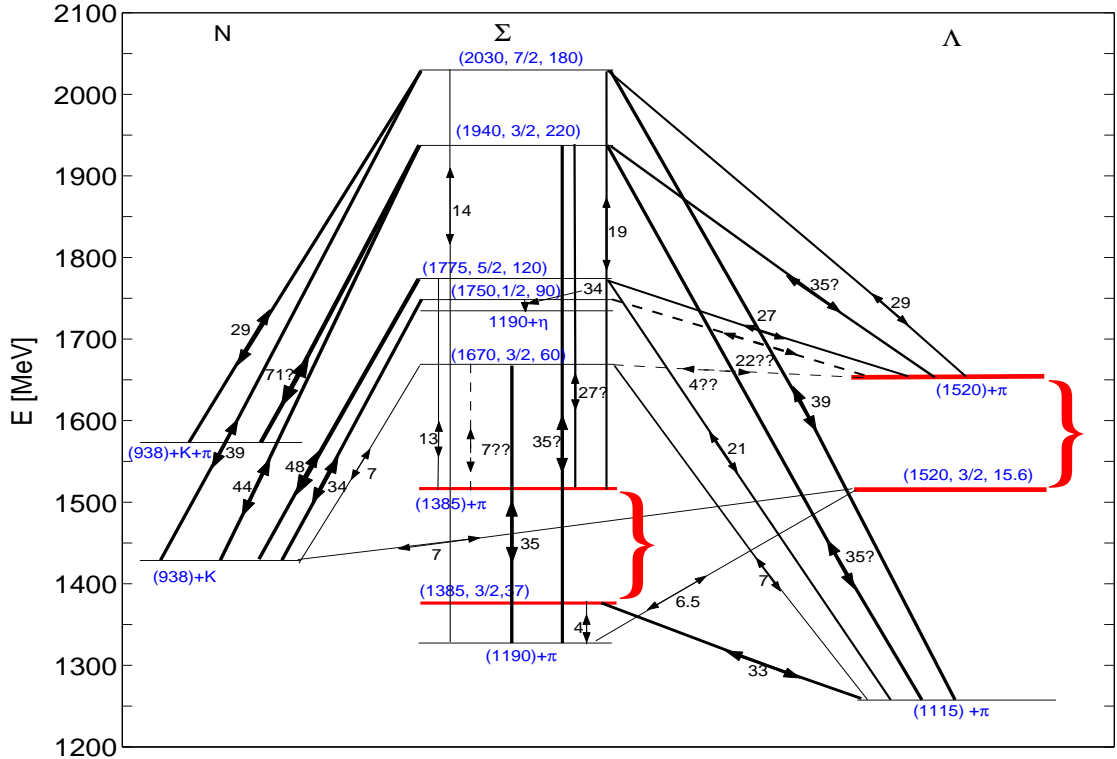}
\vskip -0.51cm
\caption{\small{ (color on line)  Reactions scheme for $\Lambda(1520)$ and
$\Sigma(1385)$  population evolutions.} } \label{Lam1520}
\end{figure*}

The resonance suppression, or enhancement, mechanism works as follows.
In thermal hadronic gas the reaction,
\begin{equation}
1 + 2 \leftrightarrow  3, \label{dpr}
\end{equation}
can occur  in both directions: the resonance decay $3\to 1+2$, and
the back-reaction (regeneration) resonance formation $1+2\to 3$.
When the reaction  goes with the same rate in both directions,  we
have chemical  detailed balance, e.g. particles yields do not change
in this period of temporal evolution of the system. This does not
necessarily mean that we have a chemical equilibrium. Instead it may
be a transient condition for which none of the three particles is
equilibrated chemically - we will show when this can happen.

In the study of resonance decay and regeneration we are using the
momentum integrated  population master equations. We assume  a
fireball  expansion model governed by hydrodynamic inspired flow
with conserved  entropy content. In our considerations we presume
that the yield of pions $\pi$ is so large that we can assume it not
to be materially affected by any of the reactions we consider. Thus
we fix pion yield in terms of an ambient fugacity and temperature
value, and in essence the total (per unit rapidity at RHIC) yield is
fixed since we conserve entropy.

An important assumption implied  below is that the rapidly expanding
hadron system maintains for the relevant particles a fully thermal
(Boltzmann) momentum distribution.
To describe the evolution of hadron abundances in the kinetic phase
we  track in time the yields of single strange hadrons after their initial formation.
This is implemented in terms of time dependence of
the chemical fugacities $\Upsilon(t)$, and  the time dependence of the
hadronization temperature $\,T(t)$.

We  look in detail at three potential evolution scenarios:\\
a) a high temperature  breakup
at $T_0\simeq 180$ MeV where the entropy content of the equilibrated QGP and HG-phase are similar;\\
b) the   $T_0\simeq 160$ MeV case where  chemical non-equilibrium among produced hadrons is already
required; and \\
c) at $T_0\simeq 140$ MeV  which is favored by descriptions of stable hadron production, and  in which case a
strong chemical non-equilibrium situation arises.

For the late stage of the expansion, at relatively low density
 the assumption of thermal momentum distribution
may not be anymore  fully satisfied. In particular pions of high
momentum could  be escaping from the fireball.  For this reason we
will consider here a second scenario, which we call ``dead
channel''.  In this scenario we assume that the reaction (\ref{dpr})
goes mainly in the direction of resonance 3 decay and the resonance
formation  is switched off for
\begin{equation}
m_3-(m_1+m_2)>300\,{\rm MeV}. \label{dchcon}
\end{equation}

Without a complete kinetic model including equilibration and
particle emission we do not know the exact energy in condition
(\ref{dchcon}) and timescale (during expansion) for which Boltzmann
distribution is violated and dead channels appear.  It is possible
that reality lies between the two cases (kinetic Boltzmann
distribution and  dead-channels) considered here which, in our
opinion, are the two most extreme limits.

For $\Lambda(1520)/\Lambda^0$ ratio calculations we employ and develop further the approach
used   for $\Sigma(1385)/\Lambda^0$ in~\cite{Kuznetsova:2008zr}.
However, in chapter ~\ref{chapter 2.1} we  investigate many further reactions in which
 resonance $\Lambda(1520)$ participates. Thus we are obliged to develop a completely
numerical evolution, for which the  analytical study of
$\Sigma(1385)/\Lambda^0$ provides a benchmark check of our approach.
In addition to new and numerous reaction channels we also introduce
deformation of the reaction rates due to stimulated Bose enhancement
of the reactions. This formalism is presented in
chapter~\ref{chapter 2.2}. We discuss the temporal evolution of HG
particle fugacities $\Upsilon(t)$ in~\ref{chapter 3.1} In
chapter~\ref{chapter 3.2} we present results for the evolution of
particle $\Sigma(1385)$, $\Lambda(1520)$ multiplicities during
kinetic phase. In chapter~\ref{chapter 3.3} we obtain the observable
`ob' ratios $\Lambda(1520)_{\rm ob}/\Lambda_{\rm tot}$ and
$\Sigma(1385)_{\rm ob}/\Lambda_{\rm tot}$.
We discuss our results in section~\ref{chapter 4}

\section{Kinetic Equations}\label{chapter 2}
\subsection{Reactions scheme for $\Lambda(1520)$ and $\Sigma(1385)$}\label{chapter 2.1}

In figure~{\ref{Lam1520}} we show the scheme of reactions  which all
have a noticeable effect on $\Lambda(1520)$ yield after the chemical
freeze-out kinetic phase. The format of this presentation is
inspired by nuclear reactions schemes. On the vertical axis the
energy scale is shown in MeV. There are three classes of particle
states, which we denote from left to right as "$N$" (S=0 baryon),
"$\Sigma$" ($S=-1, I=1$ hyperon) and "$\Lambda$" ($S=-1, I=0$
hyperon).
 Near each particle  bar   we state (on-line in blue) its mass,
and/or angular momentum and/or total width in MeV. The states
$\Lambda(1520)$ and $\Sigma(1385)$ are shown along with the location
in energy of $\Lambda(1520)+\pi$ and $\Sigma(1385)+\pi$
respectively, both entries are connected by the curly bracket, and
are highlighted (on-line in red). The  inclusion of the $\pi$-mass
is helping to see the kinetic threshold energy of a reaction. The
lines connecting the $N,\Sigma,\Lambda$ columns are indicating the
reactions we consider in the  numerical computations. All reactions
shown in figure~{\ref{Lam1520}} can go in both directions, as shown
by the double arrows placed next to the numerical value of the
partial decay width $\Gamma_i$ in MeV.

$\Lambda(1520)$ decays with a total decay width of about 15.6 MeV,
with two main channels:
\begin{eqnarray}
 \Sigma +\, \pi  \leftrightarrow \Lambda(1520) , \quad\Gamma \approx 6.5\, {\rm MeV}; \\\notag
    N + K  \leftrightarrow \Lambda(1520) ,\quad\Gamma \approx 7\phantom{.5}\, {\rm MeV}.
\end{eqnarray}
However, $\Lambda(1520)$ reacts with several
heavier $\Sigma^*$-resonances,
($\Sigma^*\equiv \Sigma(1670)$, $ \Sigma(1750)$, $ \Sigma(1775)$, $\Sigma(1940)$, $\Sigma(2030)$):
\begin{equation}
   \Lambda(1520)+\pi \leftrightarrow \Sigma^*, \label{LS*}
\end{equation}
and these reactions have a larger reaction strength
 shown  in  figure~{\ref{Lam1520}}.   $\Lambda(1520)$ nearly behaves
like a `stable' hadronic particle since:\\
a) it is dominantly coupled to heavier resonances; \\
b) its natural lifespan is  larger than the hadronic reaction rate.\\

 Hereto we note that (several) $\Sigma^*$ involved in Eq.\, (\ref{LS*}) participate   in further reactions:
\begin{eqnarray}
&&   \Lambda(1115) +\pi  \leftrightarrow \Sigma^*; \label{L1115S*}\\   
&&   \Sigma(1190)+\pi \leftrightarrow  \Sigma^*;  \\  
&&    N + K  \leftrightarrow \Sigma^*;  \\
&&    \Sigma(1385) + \pi  \leftrightarrow \Sigma^*; \label{S1385S*}\\ 
&&   \Delta + K  \leftrightarrow \Sigma(1940, 2030) ;    \\ 
&&   N + K(892)  \leftrightarrow \Sigma(1940);   \\ 
&&   \Sigma   + \eta\leftrightarrow  \Sigma(1750).
\end{eqnarray}
All  reactions shown above can excite $\Sigma^*$ resonances. Since
the mass of $\Lambda(1520)$ is near to the
$\Sigma^*$ mass, the yield of  $\Lambda(1520)$ is effectively depleted
by the   reaction chain
\begin{equation}
\Lambda(1520)+\pi  \rightarrow  \Sigma^* \rightarrow N+{\rm K},\  {\rm etc} .\label{LS*N}
\end{equation}
The balancing
two step back-reaction can also occur, especially once $\Lambda(1520)$ has been  depopulated.
Thus   a dynamical reduced   detailed balance yield of  $\Lambda(1520)$ would result
if the system were at fixed volume rather than  expanding.

The multiplicity of $\Sigma(1385)$ is mostly determined by its dominant
decay and production in the reaction
\begin{equation}
  \Lambda(1115) + \pi  \leftrightarrow \Sigma(1385)  , \label{L1115L}
\end{equation}
and to a lesser extent by  the reaction
\begin{equation}
  \Sigma(1190) + \pi \leftrightarrow \Sigma(1385)  . \label{S1385S}
\end{equation}
The resonance $\Sigma(1385)$ participates further in reactions with
heavier $\Sigma^*$; see reaction (\ref{S1385S*}), but strength of
these interactions is smaller than for similar reactions with
$\Lambda(1520)$ and smaller than the decay width of $\Sigma(1385)$.
Thus we find that the  influence of these reactions on
$\Sigma(1385)$ yield is small. Another reason for a reduced
effective depletion rate of  $\Sigma(1385)$ is that a lesser
fraction of this resonance is needed to excite $\Sigma^*$. Thus in
such a reaction the depopulation effect decreases  because of a
larger mass difference between $\Sigma(1385)$ and $\Sigma^*$ in
comparison with  $\Lambda(1520)$ and $\Sigma^*$.

\begin{figure*}
\centering
\includegraphics[width=17 cm, height=12 cm]{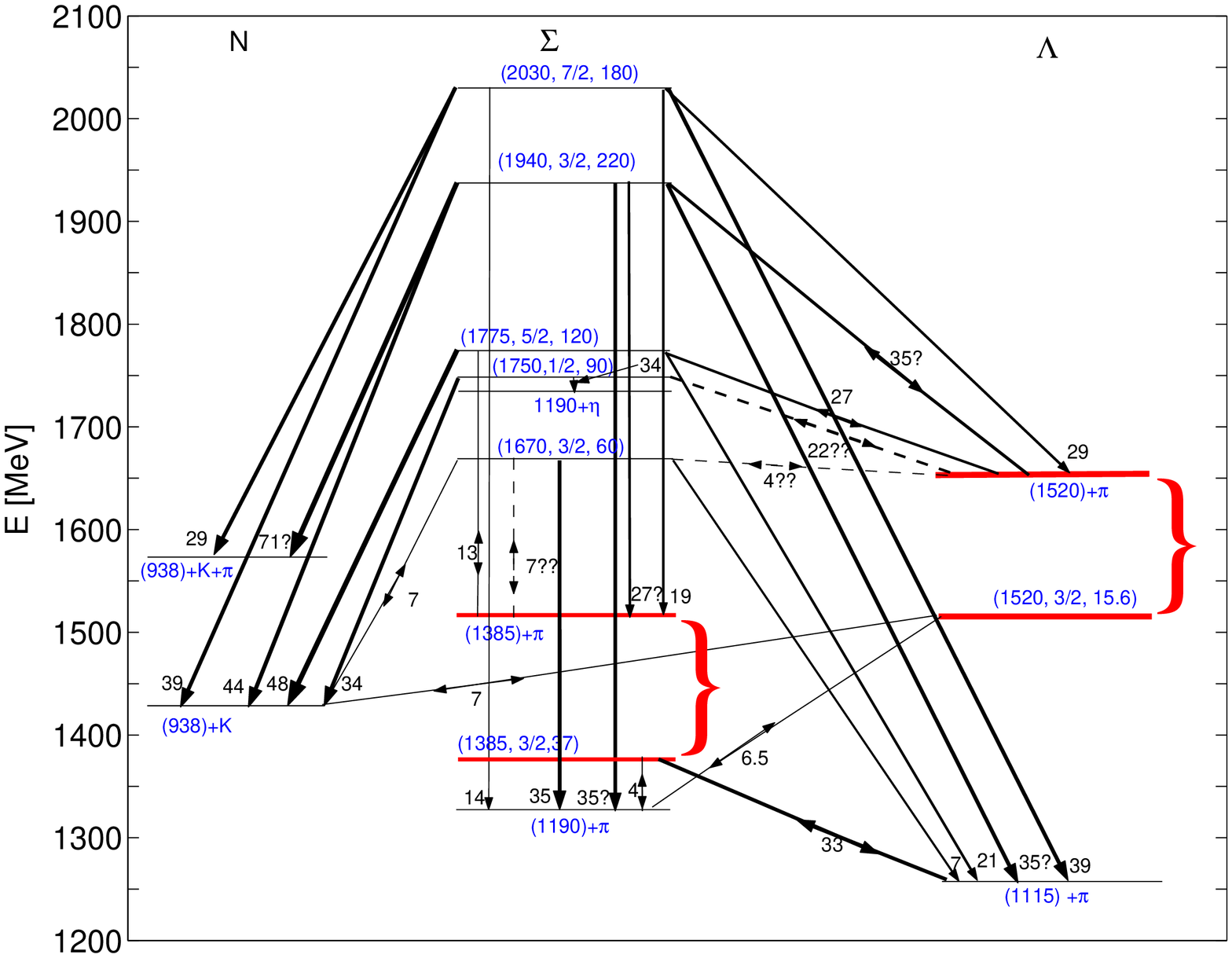}
\vskip -0.51cm \caption{\small{ (color on line)  Reactions scheme
for $\Lambda(1520)$ and $\Sigma(1385)$ interactions in the ``dead
channel'' model.} } \label{Lam1520dc}
\end{figure*}

The reactions scheme for $\Lambda(1520)$ reactions with dead
channels is shown in figure~\ref{Lam1520dc}. The difference between
figure~\ref{Lam1520}  and figure~\ref{Lam1520dc} is that some of the
reaction lines have single-directional arrows, as is stipulated by
the condition Eq.\,(\ref{dchcon}).

\subsection{Resonances densities, time evolution equations}\label{chapter 2.2}

The  evolution in time  of the resonance yield is described by a
master equation, where the process of resonance formation in
scattering is balanced by the natural resonance decay:
\begin{equation} \label{delev}
\frac{1}{V}\frac{dN_{3}}{dt}=\sum_i\frac{dW^i_{{1+2 \rightarrow
3}}}{dVdt}-\sum_j\frac{dW^j_{{3 \rightarrow 1 +2}}}{dVdt},
\end{equation}
where subscripts $i$, $j$ denote different reactions channels when
available. We further allow different subscripts $i$, $j$ for the
case where there are dead channels. Thus ${dW^i_{1+2 \rightarrow
3}}/{dVdt}$ and ${dW^j_{{3 \rightarrow 1 + 2}}}/{dVdt}$ are
invariant rates (per unit volume and time) for particle $3$
production and decay respectively.   In case all reactions occur in
both directions the total number of fusion channels is the same as
the total number of decay channels.

Allowing for Fermi-blocking and Bose enhancement in the final state,
where  by designation   particles $1$  and $3$ are fermions (heavy baryons) and
particle $2$ is a boson (often light pion) we have for the two rates:
\begin{eqnarray}
\frac{dW^j_{3 \rightarrow 1 +2 }}{dVdt}=
&&  \int\frac{g_{3}d^{3}p_{3}}{2E_{3}(2\pi)^3}f_{3}
     \int\frac{d^{3}p_{1}}{2E_1(2\pi)^3}\left(1 - f_{1}\right)\nonumber\\[0.3cm]
&&\hspace*{-1.4cm}
\times\int\frac{d^{3}p_{2}}{2E_{2}\left(2\pi\right)^{3}}\left(1 + f_{2}\right)
\left(2\pi\right)^{4}\delta^{4}\left(p_{1}+p_{2}-p_{3}\right)\nonumber\\[0.3cm]
&&\hspace*{-.4cm}
\times\frac{1}{g_{3}}\sum_{\rm spin}\left|\langle p_{3}\left|
M^j\right|p_{1}p_{2}\rangle\right|^{2}\label{dr},
\end{eqnarray}
\begin{eqnarray}
\frac{dW^i_{1+2 \rightarrow 3}}{dVdt}&=&
\int\frac{g_1d^{3}p_{1}}{2E_{1}(2\pi)^3}f_{1}
\int\frac{g_{2}d^{3}p_{2}}{ 2E_{2}(2\pi)^3}f_{2}  \nonumber\\
&&\hspace*{-1.4cm}
\times \int\frac{d^{3}p_{3}}{2E_{3}\left(2\pi\right)^{3}}\left(1 - f_{3}\right)
     \left(2\pi\right)^{4}\delta^{4}\left(p_{1}+p_{2}-p_{3}\right)\nonumber\\
&&\hspace*{-.4cm}
\times\frac{1}{g_1g_{2}}\sum_{\rm{spin}}\left|
\langle p_{1}p_{2}\left| M^i\right|p_{3}\rangle\right|^{2}.
\label{pr}
\end{eqnarray}
where $g_i, i=1,2,3$ is particles degeneracy. The Bose distribution
function for particle $2$ is
\begin{eqnarray}
f_{2}&=&\frac{1}{\Upsilon_{2}^{-1}e^{u\cdot p_{2}/T} - 1},\label{fpi}
\end{eqnarray}
and Fermi for particles $1,3$ are:
\begin{eqnarray}
f_{j}&=&\frac{1}{\Upsilon_j^{-1}e^{u\cdot p_j/T} + 1},\ j=1,3\label{fN}.
\end{eqnarray}
Here $\Upsilon_i$ is particles fugacity, and  $u\cdot p_i=E_i$, for
$u^\mu=(1,\vec 0)$ in the rest frame of the heat bath where
 $d^4p\delta_0(p_i^2-m_i^2)\to d^3p_{i}/E_{i}$ for each particle. Hence,
Eq.(\ref{dr}) and Eq.(\ref{pr}) are  Lorentz invariant, and thus as presented these rates
can be evaluated  in any convenient frame of reference. Normally, this is the
frame co-moving with  the thermal volume element.

For the  heavy baryon  (resonances), particles  2,3, we can work using the expansion of the
relativistic  distribution, the first term is the Boltzmann limit:
\begin{eqnarray}
\frac {N_{i} }V = \Upsilon_{i}\frac{T^3}{2\pi^2}g_{i}x_{i}^2K_2(x_{i}), \label{relboltz}
\end{eqnarray}
where $x_{i}=m_{i}/T$, $K_2(x)$ is the Bessel function (not to be
mixed up with particle 2) . However, we use the complete Bose
distribution to describe pions.

We introduce in medium lifespan of particle 3:
\begin{equation}
\frac{1}{\tau_3}\equiv   \frac{\sum_i R^i_{123}}{V^{-1}dN_3/d\Upsilon_3},\label{Dect}
\end{equation}
and, similarly, channel lifespan $\tau_3^i$, omitting the sum $\sum_i$.
Here the rate $R_{123}$ is:
\begin{eqnarray}
&&R_{123}^i=
\int\!\! \!\!\int\!\! \!\!  \int\!\!
 \frac{d^3p_1d^3p_2d^3p_3}
         {8E_1E_2E_3(2\pi)^5}
f_1\Upsilon_1^{-1}f_2 \Upsilon_2^{-1} f_3 \Upsilon_3^{-1} \nonumber\\
&&\hspace*{-.4cm}
\times
    \delta^{4}\left(p_{1}+p_{2}-p_{3}\right)e^{u\cdot p_3/T}
\sum_{\rm{spin}}\left| \langle p_{1}p_{2}\left| M^i\right|p_{3}\rangle\right|^{2}.
\label{prR}
\end{eqnarray}
$R$  is independent of the fugacity in the Boltzmann-limit. In next
section \ref{chapter 2.3} we will see in what way the in-medium
decay rate varies from the free space decay rate. This is due to the
effect of quantum enhancement, and  the fact that a particle emerged
in a thermal bath with a finite temperature. A particle is not
decaying in its rest frame.

The production and decay rates are connected to each other by the
detailed balance relation~\cite{KuznKodRafl:2008,Kuznetsova:2008jt}:
\begin{equation}
\Upsilon_1^{-1} \Upsilon_2^{-1}\frac{dW_{1+2 \rightarrow 3}}{dVdt}=
 \Upsilon_{3}^{-1} \frac{dW_{3 \rightarrow 1+2}}{dVdt}=R_{123}.\label{pdr}
\end{equation}
Using detailed balance Eq.\,(\ref{pdr}) we obtain  for fugacity
$\Upsilon_3$ the  evolution
equation~\cite{KuznKodRafl:2008,Kuznetsova:2008jt}:
\begin{equation}
\frac{d\Upsilon_{3}}{d{\tau}} =
\sum_i{\Upsilon^i_{1}\Upsilon^i_{2}}\frac{1}{\tau^i_{3}}
+\Upsilon_{3}\left(\frac{1}{\tau_T}+\frac{1}{\tau_S}-\sum_j\frac{1}{\tau^j_3}\right),
\label{Ups2}
\end{equation}
where we have also introduced  characteristic time constants of temperature $T$ and entropy $S$ evolution
\begin{eqnarray}
\frac{1}{\tau_{T} } &=&  -\frac{d\ln({x_{3}}^2K_2(x_{3}))}{dT} \dot{T}, \label{Teq}\\
\frac{1}{\tau_{S} } &=&  -\frac{d\ln( VT^3  )}{dT} \dot{T}. \label{Seq}
\end{eqnarray}
The entropy term is negligible, $\tau_S\gg \tau_3, \tau_T$ since we
implement near conservation of entropy. We implement this in the way
which would be exact for massless particles taking  $VT^3=$Const..
Thus there is some entropy growth in HG evolution to consider, but
it is not significant. In order to evaluate  the magnitude of
$\tau_T$ we use the relation between Bessel functions of order 1 and
2 (not to be mixed up with particles 1,2)
${d}\left(z^2K_2(z)\right)/{dz}=-z^2K_1(z)$. We obtain
\begin{equation}
\frac{1}{\tau_{T}} = - \frac{K_1(x_{3})}{K_2(x_3)}x_3 \frac{\dot{T}}{T}, \label{Teq2}\\
\end{equation}
$\tau_{T}>0$. We  invoke a model of matter expansion
of the type used e.g. in~\cite{Letessier:2006wn}, where the longitudinal
and transverse expansion is considered to be (nearly) independent.
In this model we have:
\begin{equation}\label{DTT}
\frac{\dot {T}}{T} = -\frac{1}{3}\left( \frac{2\,(v\tau/R_{\perp}) + 1}{\tau}\right),
\end{equation}
where $R_{\perp}$ is the transverse radius, $v$ is the velocity of
expansion in the transverse dimension. All flow parameters (or
temperature dependence on $\tau$) are the same as
in~\cite{Kuznetsova:2008zr}. For  a static system with $\tau_T \to
0$ we see that Eq.\,(\ref{Ups2}) has  transient stable population
points whenever
 \begin{equation}
\sum_i\Upsilon_1^i\Upsilon_2^i\frac{1}{\tau_3^i}-\Upsilon_3\sum_j\frac{1}{\tau_3^j}=0.
\label{stable}
 \end{equation}

Next we address the functional dependence on time of
$\Upsilon_{1},\Upsilon_{2}$.   In the equation for $\Upsilon_{1}$ we
have terms which compensate what is lost/gained in  $\Upsilon_{3}$
see Eq.\,(\ref{Ups2}). Further we have to allow that particle `1'
itself plays the role of particle 3 (for example this is clearly the
case for $\Lambda(1520)$). That allows a chain of populations
relations as follows:
 \begin{equation}
(1'+2' \leftrightarrow 1) + 2 \leftrightarrow  3, \label{dpr2}
 \end{equation}
Then we  obtain:
\begin{eqnarray}
\frac{d\Upsilon_{1}}{d{\tau}}\!\! &=& \!\!
\Upsilon_{3}\sum_k\frac{1}{\tau^k_3}\frac{dN^k_3/d{\Upsilon^k_3}}{dN_1/d{\Upsilon_1}}
-\sum_n{\Upsilon_{1}\Upsilon^n_{2}}\frac{1}{\tau^n_{3}}\frac{dN^n_3/d{\Upsilon^n_3}}{dN_1/d{\Upsilon_1}}\nonumber\\
&+& \Upsilon_{1}\left(\frac{1}{\tau_T}+\frac{1}{\tau_S}-\sum_j\frac{1}{\tau^j_1}\right)
+\sum_i{\Upsilon^i_{1'}\Upsilon^i_{2'}}\frac{1}{\tau^i_{1}}
\label{Ups3}
\end{eqnarray}
The ratios of derivative of $N_i$ seen in the first line are due to
the definition of relaxation time Eq.\,(\ref{Dect}). The system  of
equations for baryons closes with the equation for $\Upsilon_{1'}$
 \begin{eqnarray}
 \frac{d\Upsilon_{1'}}{d{\tau}} \!\! &=  &\!\!
\Upsilon_{1}\sum_k\frac{1}{\tau^k_1}\frac{dN^k_1/d{\Upsilon^k_1}}{dN_{1'}/d{\Upsilon_{1'}}}
-\sum_n{\Upsilon_{1'}\Upsilon^n_{2'}}\frac{1}{\tau^n_{1}}\frac{dN^n_1/d{\Upsilon^n_1}}{dN_{1'}/d{\Upsilon_{1'}}}\nonumber\\
&+&  \Upsilon_{1'}\left(\frac{1}{\tau_T}+\frac{1}{\tau_S}\right).
\label{ups4}
 \end{eqnarray}
In the present setting  $\Upsilon_{2=\pi}=$Const. by virtue of
entropy conservation (see discussion below) and the same applies to
the case $2'=\pi$. However, if either particle $2$ or  $2'$ is a
kaon, we need to follow the equation for $\Upsilon_{2,2'=K}$ which
is analogous  to equation for particle $1$ or $1'$.\

The evolution equations can be integrated once we determine the {\em
initial} values of particle densities (fugacities) established at
hadronization/chemical freeze-out. We determine these for RHIC
head-on Au--Au collisions at $\sqrt{s_{\rm NN}}=200$ GeV. We
introduce  the  initial hadron yields  inspired by a picture of a
rapid hadronization of QGP in which quarks combine into final state
hadrons. For simplicity we assume here that  the net baryon yield at
central rapidity  is negligible. Thus the baryon-chemical and
strangeness potentials vanish. The initial yields  of mesons ($q\bar
q, s\bar q$) and baryons ($qqq, qqs$) are controlled aside of the
ambient temperature $T$ by the  constituent light quark fugacity
$\gamma_q$ and the strange quark fugacity~$\gamma_s$.

The strangeness pair-yield in QGP is maintained in transition to HG.
This fixes the initial value of $\gamma_s$. In fact, since we
investigate here relative chemical equilibrium reactions our results
do not depend significantly  on the exact initial value $\gamma_s$
 and/or strangeness content. The entropy
conservation at hadronization fixes $\gamma_q$. For hadronization
temperature $T(t=0)\equiv T_0=180$ MeV,  $\gamma_q=1$. However, when
$T_0<180$ MeV, $\gamma_q>1$ in order to have entropy conserved at
chemical freeze-out. At $T_0=140$ MeV $\gamma_q=1.6$ that is close
to maximum possible value of $\gamma_q$, defined by Bose-Einstein
condensation condition \cite{Kuznetsova:2006bh}.

For reactions, such as  shown in Eq.\, (\ref{dpr}), we have (lower
index defines particle considered, where $Y\equiv \Sigma, \Lambda$
is a hyperon)
\begin{equation}
\Upsilon^0_{(1=Y)}=\gamma_q^{2}\gamma_s,\qquad 
\Upsilon^0_{(2=\pi)}=\gamma_q^{2}; \qquad \label{upinpi}
\end{equation}
or
\begin{equation}
\Upsilon^0_{(1=N)}=\gamma_q^{3}, \qquad
\Upsilon^0_{(2={\rm K)}}=\gamma_q\gamma_s; \qquad \label{upinpi1}
\end{equation}
where the particle 1 in reaction (\ref{dpr}) is a baryon and particle 2
is a meson. The particle 3 is always a strange baryon:
\begin{equation}
\Upsilon^0_{(3=Y)}=\gamma_q^{2}\gamma_s, 
\end{equation}
Note that  for  $\gamma_q > 1$  we have always initially
\begin{equation}
\left.\frac{\Upsilon_1
\Upsilon_2}{\Upsilon_3}\right\vert_{t=0}=\gamma_q^2 \ge 1\,.
\end{equation}
As a consequence initially the pair of particles 1,2 reacts into 3.

As already noted, we do not need to follow the evolution in time for
the pion yield, which is fixed by conservation of entropy per unit
rapidity, as incorporated in Eq.\,(\ref{DTT}). Thus it is
(approximately) a constant of motion. This can be seen recalling
that the  entropy per pion is nearly 4 within the domain of
temperatures considered. Thus the conservation of entropy implies
that pion number is conserved. With $VT^3\simeq \mathrm{Const.}$,
this further implies that during the expansion
$$\Upsilon_\pi=\gamma_q^2=\mathrm{Const.},$$
which we keep at the initial value.

\subsection{In medium lifespan calculations}\label{chapter 2.3}

In our calculations we take into account the influence of the medium
on resonance lifespan and the effect of the motion of the decaying
particle with respect to the thermal rest frame.
In~\cite{KuznKodRafl:2008,Kuznetsova:2008jt} it was noted that the
decay rate $R_{123}$, Eq.\,(\ref{prR}), of particle $3$  (density)
in a thermally equilibrated system can be cast  into a form which
involves the free space decay rate:
\begin{equation}
R_{123}=
\frac{m_{3}}{\tau _{0}}\int_{0}^{\infty }\frac{p_{3}^{2}dp_{3}}{E_{3}}
\frac{
\Upsilon_{3}^{-1}
e^{E_{3}/T}}{\Upsilon_{3}^{-1}e^{E_{3}/T}\pm 1}\Phi (p_{3}),  \label{Decay1-final}
\end{equation}
where function $\Phi (p_{3})$ for reaction (\ref{L1115L}) $\Sigma(1385)\leftrightarrow \Lambda+\pi$ is
\begin{equation}
\Phi(p_3)=\frac{1}{b(e^{E_{3}/T} \!+\! \Upsilon_{\pi}\Upsilon_{\Lambda})}
\ln\frac{\!\left(\Upsilon_{\Lambda}e^b+e^{-a_2} \right)
\!\left(e^{a_1} \!- \! \Upsilon_{\pi}e^{-b}\right)}
{\!\left(\Upsilon_{\Lambda}e^{-b}+e^{-a_2}\right)
\!\left(e^{a_{1}} \!- \!\Upsilon_{\pi}e^b\right)}.\notag
\label{phia1a2bf}
\end{equation}
\begin{equation}
a_{1}=\frac{E^{*}_{1}E_{3}}{m_{3}T}, \quad
a_{2} =\frac{E^{*}_{2}E_{3}}{m_{3}T}, \quad
b =\frac{p^{*}p_{3}}{m_{3}T}.
\end{equation}
Here $p^{*}=p_1=p_2$ and $E^{*}_{1,2} =\sqrt{p^{*\,2}+m^2_{1,2}}$
are the magnitude of the momentum and, respectively, the energy, of particles $1$ and $2$ in the rest frame of the particle $3$.
From energy conservation:
\begin{eqnarray}
 E^{*}_{1,2}&=&\frac{m_{3}^{2}\pm (m_{1}^{2}-m_{2}^{2})}{2m_{3}}, \nonumber\\
 p^{*\,2} &=& E_{1,2}^{2}-m_{1,2}^{2}   \nonumber\\
&=&\frac{m_3^2}{4}-\frac{m_1^2+m_2^2}{2}+\frac{(m_1^2-m_2^2)^2}{4m_3^2}.
\label{encon}
\end{eqnarray}

\begin{figure}[tbp]
\centering \includegraphics[width=8.5cm,height=8.5cm]{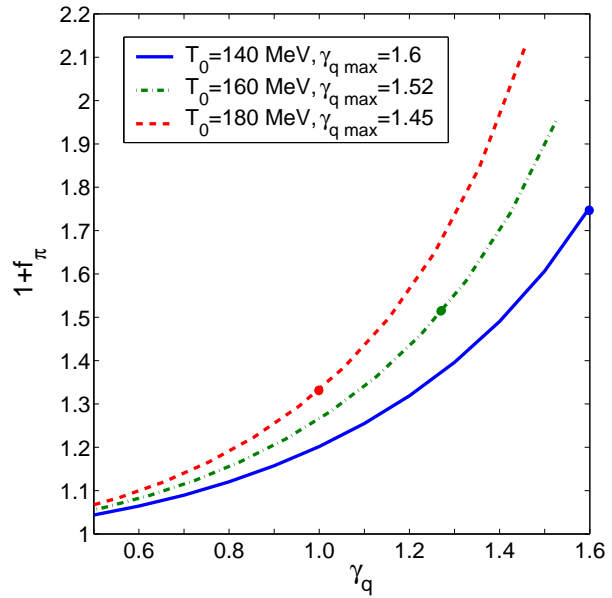}
\vskip -0.31cm
\caption{\small (color on line) The Bose enhancement factor $1+f_{\pi}(E^{*}_1)$
in $\Sigma(1385)$ rest frame
as a function of light quark fugacity $\gamma_q$ for the
reaction $\Sigma(1385) \leftrightarrow \Lambda \pi$ at $T=140$ MeV (blue, solid line),
at $160$ MeV (green, dash-dot line) and $180$ MeV (red, dashed line). The dots
show  the initial value of  fugacities  for the three possible
hadronization cases.} \label{bose140}
\end{figure}

Here fugacities for $\Lambda$ and $\pi$  correspond to those for
particles 1 and 2, respectively. For the temperatures of interest
(hadronization of QGP and below) $m_{\Lambda}$ and $m_{\Sigma} >>
T$. With sufficient accuracy we can write
\begin{equation}
\Phi(p_3)\simeq \frac{1}{be^{E_{3}/T}}\ln \frac{\left(e^{a_1+b} - \Upsilon_{\pi}\right)}{\left(e^{a_1-b} -\Upsilon_{\pi}\right)}.
\label{phia1a2bfnr}
\end{equation}
There are no significant medium  effects upon decay rate of
$\Sigma(1385)$ and $\Lambda$  resonances. However the pions have
energy $E^{*}_2 = 250$ MeV (Eq.(\ref{encon})) in the $\Sigma$ rest
frame and the Bose enhancement effect is possible in the
oversaturated   hadronic gas after QGP hadronization.

For the  low temperatures considered here we can assume that
$\Sigma$ resonances almost do not move.  Thus the enhancement effect
in the thermal bath  frame is close to the enhancement in the
$\Sigma(1385)$ rest frame. The decay  rate increases by Bose
enhancement factor $1+f_{\pi}$ (here $f_{\pi}=f_{\pi}(E^{*}_2, T)$).
In figure \ref{bose140} we show Bose enhancement factor as a
function of light quark fugacity $\gamma_q$ for temperature
$T_0=140$ MeV (blue, solid line), $T_0=160$ MeV (green, dash-dot
line), $T_0=180$ MeV (red, dashed line). The large dots show Bose
enhancement factor for our initial $\gamma_q$ determined from
entropy conservation in  fast hadronization. The fugacity
$\gamma_q=1.6$ is close to maximum expected value at $T_0=140$ MeV.
The maximum fugacities for each temperature correspond to Bose -
Einstein singularity. The Bose enhancement effect is largest for
maximum $\gamma_q$ and it diminishes  for small $\gamma_q$. At fixed
entropy the greatest enhancement is for smallest ambient
temperature, see the dot on solid line in figure \ref{bose140}.

\begin{figure}
\centering \includegraphics[width=9.cm,height=9.cm]{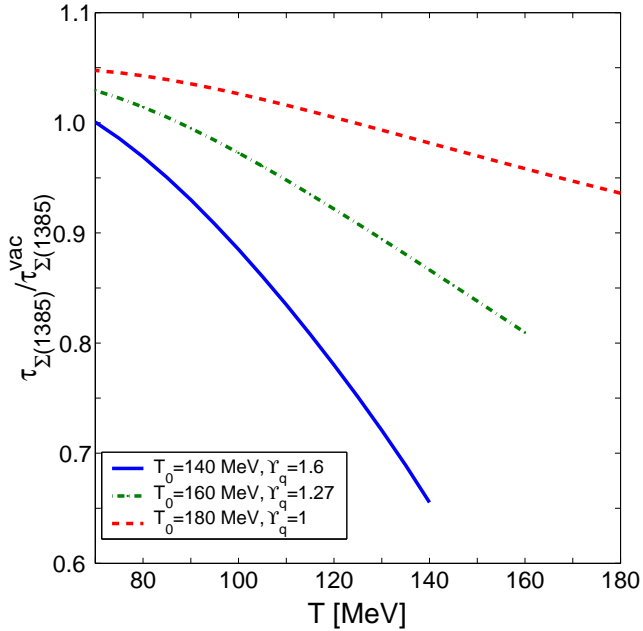}
\vskip -0.31cm
\caption{\small    (color on line)
The ratio  of the in medium lifespan  $\protect\tau_{3}$  with the vacuum
lifespan $\tau _{0}$ as a function of temperature $T$ for the reaction
$\Sigma(1385)\leftrightarrow \Lambda\pi$. The  dashed (red) line is for
hadronization at $T_0=180$ MeV, $\gamma_q=1.0$; the dot-dashed line (green)
for hadronization at $160$ MeV, $\gamma_q=1.27$; solid line (blue) is for
hadronization at $140$ MeV and   $\gamma_{q}= 1.6$.}\label{tausig}
\end{figure}

\begin{figure*}
\centering
\hspace*{-.3cm}\includegraphics[width=18.5 cm]{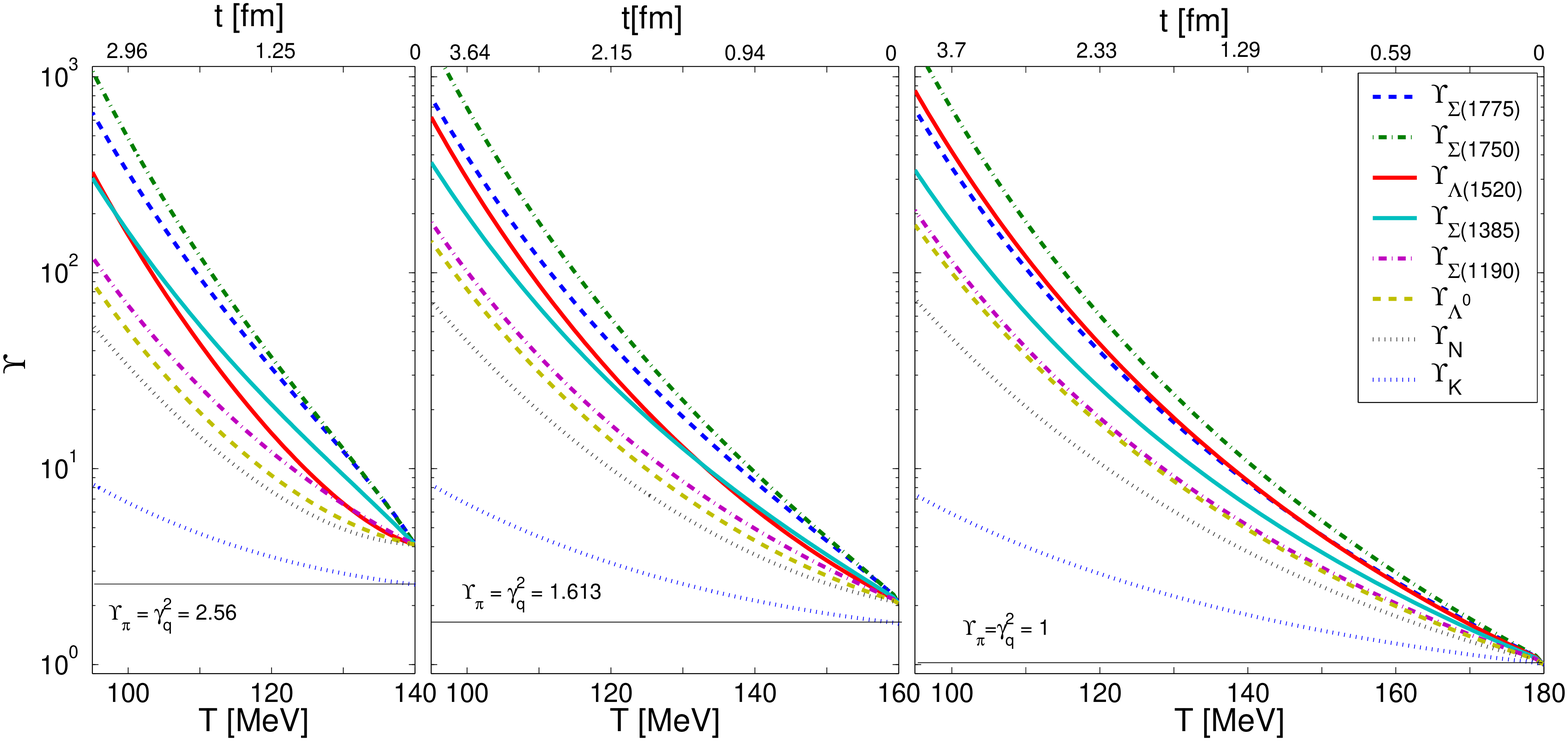}
\caption{\small{ (color on line) The fugacities $\Upsilon$  for selected particles are shown as a function of
temperature $T(t)$, for $T_0=140$ MeV on the left, for $T_0=160$ MeV in the middle and
for $T_0=180$ MeV, on the right. See text for further details.}} \label{Upsil}
\end{figure*}

In figure \ref{tausig} we show the corresponding decrease in the
lifespan, the ratio $\protect\tau_{3}/\protect\tau _{0}$ as a
function of temperature $T$ in the reaction
$\Sigma(1385)\leftrightarrow \Lambda\pi$. We consider temperature
range from corresponding hadronization temperature until $T=70$ MeV.
We assumed, that $\Upsilon_{\pi}$ is a constant. Fugacities of heavy
resonances do not influence the result. The lowest $\tau_3/\tau_0$
ratio is for $\gamma_q=1.6$ at $T_0=140$ MeV when we have maximum
value of $\gamma_q$ for given temperature. If we compare this value
of $\tau_3/\tau_0 = 0.65$ with inverse Bose enhancement factor
$1/(1+f_{\pi}(E_2^*, T)) = 0.54$ for this $T$ and $\gamma_q$ (see
figure \ref{bose140}) we see that these values are near to each
other (difference is about 20\% ) as expected for $m_{\Sigma}>>T$.
For smaller $T$, $\gamma_q$ decay time goes to its vacuum value.

The same calculations are applicable for heavier $\Sigma^*$.  When
the difference of mass of the initial and final state resonance
decreases, the Bose enhancement effect increases, since it involves
small momenta. The largest effect is for reaction $\Sigma(1670)
\leftrightarrow \Lambda(1520) + \pi$. On the other hand, for the
reactions which satisfy condition (\ref{dchcon}) the enhancement
effect becomes so small that we do not need to include it in our
calculations.

\section{Numerical results}\label{chapter 3}
\subsection{Evolution of fugacities}\label{chapter 3.1}

In order to evaluate the $\Lambda(1520)$ and $\Sigma(1385)$
multiplicities we must  integrate  Eq.\,(\ref{Ups2}), or
Eq.\,(\ref{Ups3}), or Eq.\,(\ref{ups4}) for each particle involved
in figure~\ref{Lam1520}, and perform similar operations for
reactions with dead channels in figure~\ref{Lam1520dc}. This system
of equations includes equations for $\Lambda(1520)$, $\Sigma(1385)$,
five equations for $\Sigma^*$s, equations for K(892) and $\Delta$
and equations for ground states $\Lambda(1115)$, $\Sigma(1190)$, N,
K. All reactions in figures~{\ref{Lam1520}} are included. We solve
this system of equations numerically, using classical fourth order
Runge-Kutta method.

\begin{figure*}
\centering
\includegraphics[width=8 cm, height=10 cm]{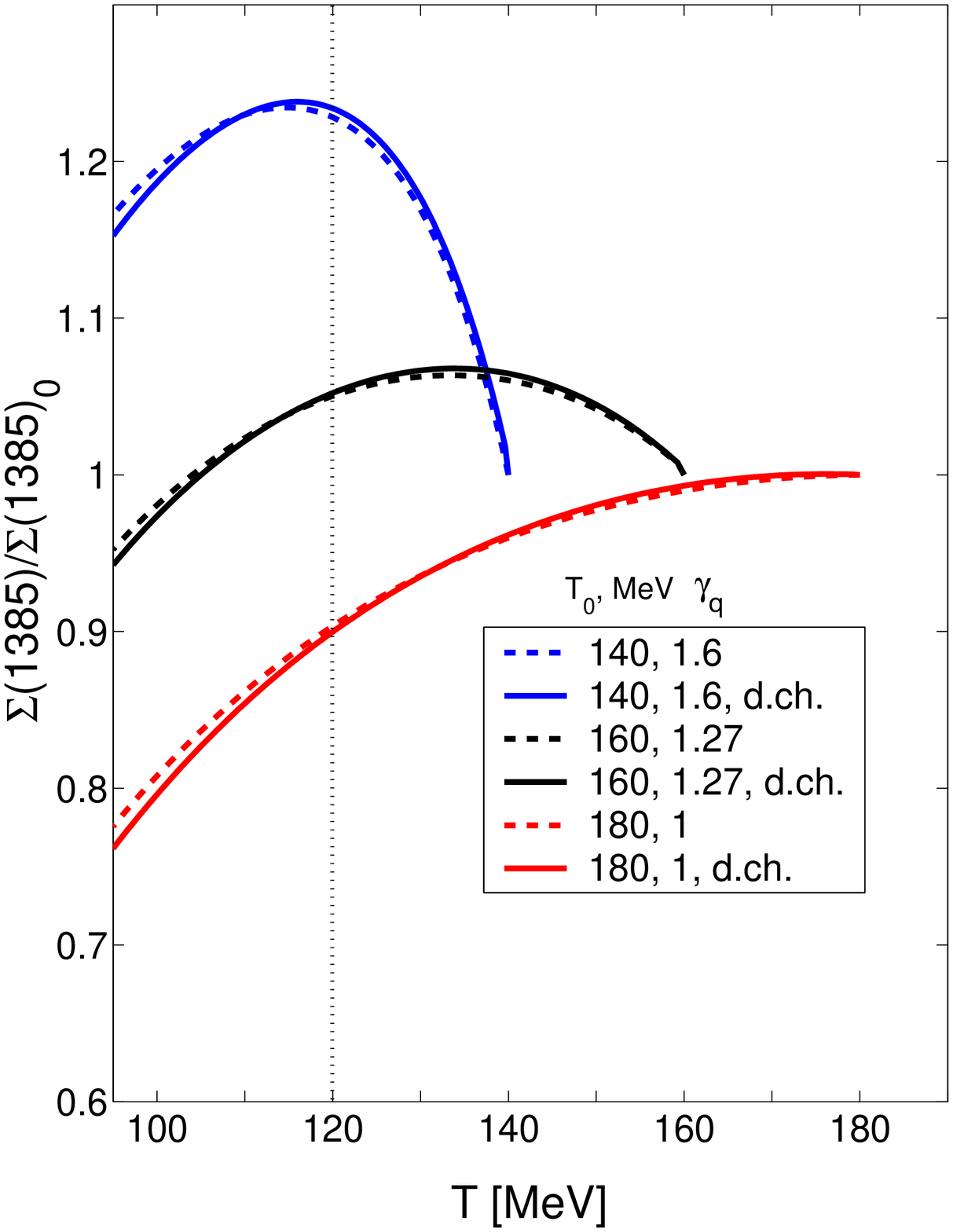}
\includegraphics[width=8 cm, height=10 cm]{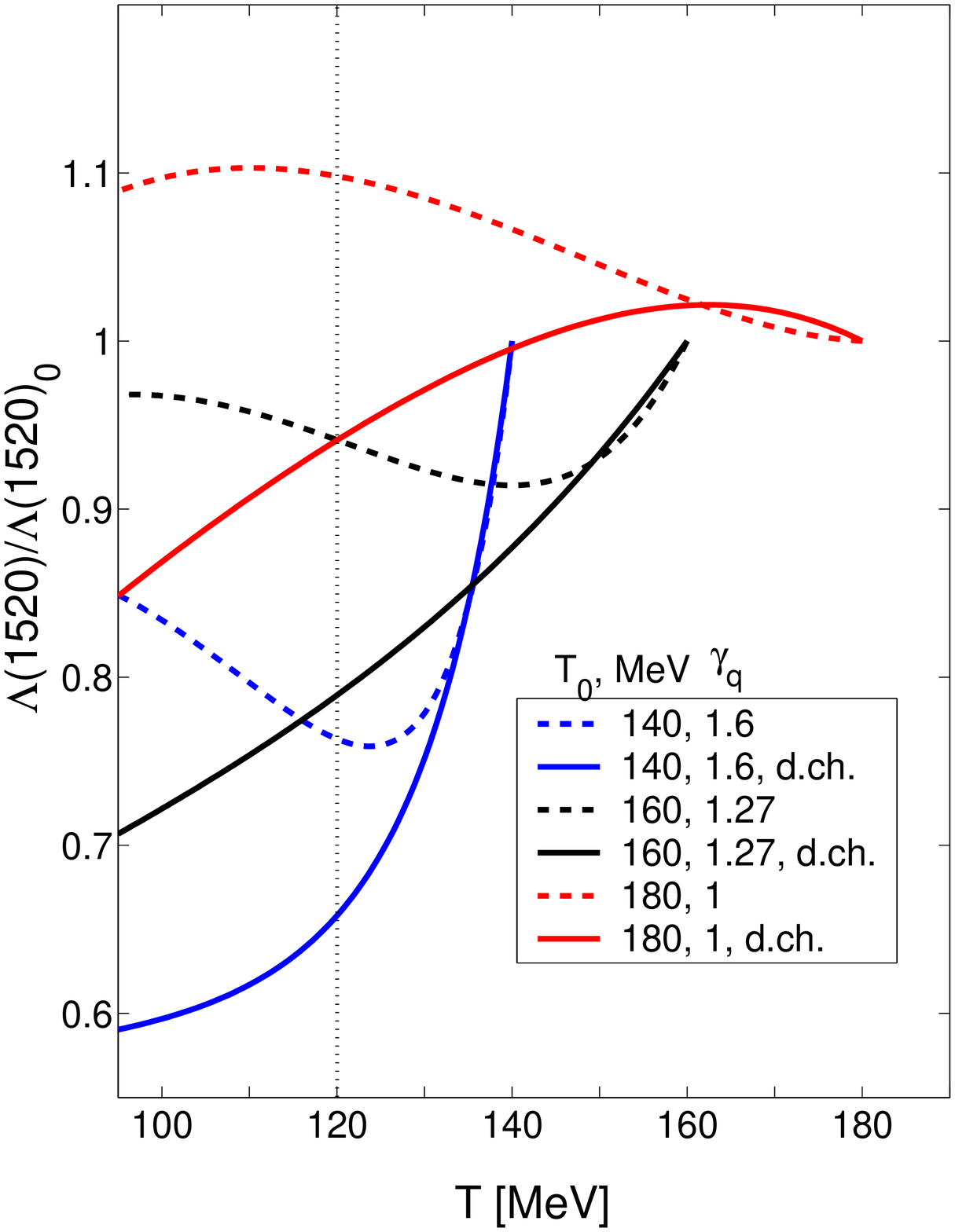}
\caption{\small{The ratio
$\Sigma(1385)/\Sigma(1385)_0$ on left
and $\Lambda(1520)/\Lambda(1520)_0$ on right
as a functions of temperature $T(t)$ for
different initial hadronization temperatures $T_0=140$, $160$
and $180$ MeV (blue/bottom, black/middle and red/top lines, respectively). Solid lines
are for calculations with dead channels, dashed lines are for
calculations without dead channels.}} \label{Lam1520r}
\end{figure*}

\begin{figure*}
\centering
\includegraphics[width=8 cm, height=10 cm]{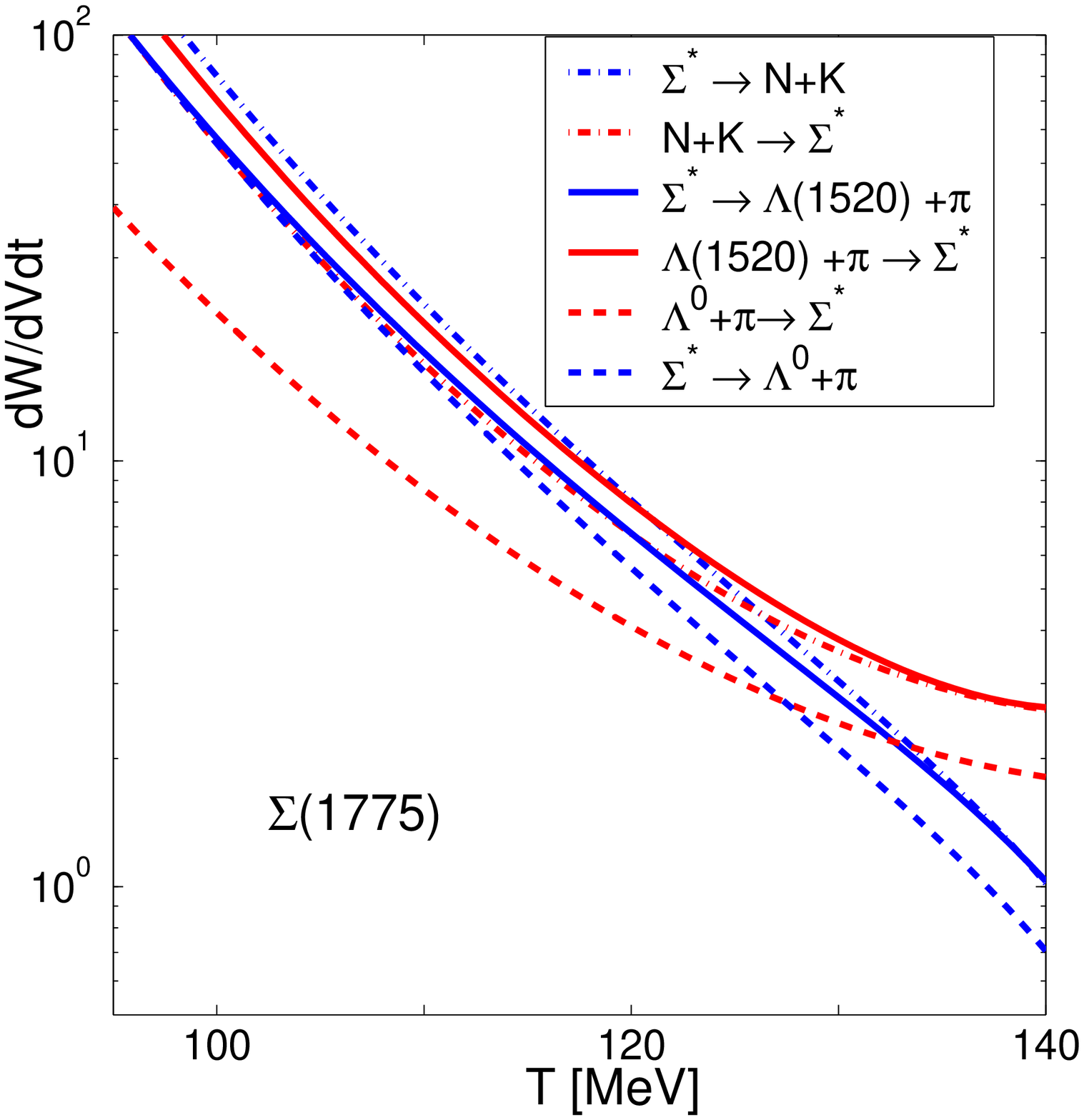}
\includegraphics[width=8 cm, height=10 cm]{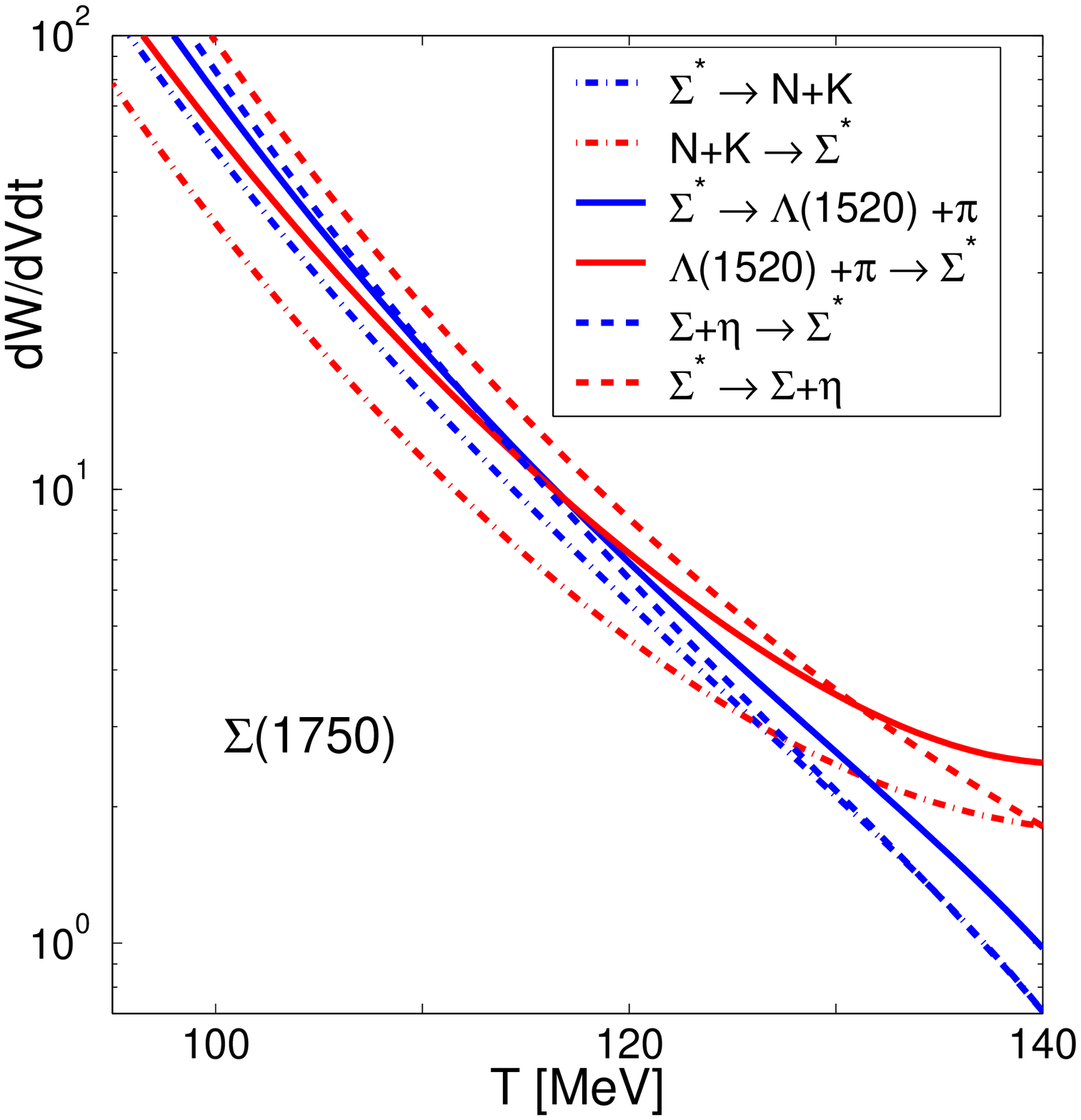}
\caption{\small{The rates for main channels of $\Sigma(1775)$ (on
the left) and  $\Sigma(1750)$ (on the right) decay and production as
a functions of temperature $T$  in the case when all reactions go in
both directions and $T_0=140$ MeV. Solid lines  are for reaction
$\Sigma^* \leftrightarrow \Lambda(1520) + \pi$; dash-dot lines are
for reaction $\Sigma^* \leftrightarrow N+K$; dashed lines are for
reaction $\Sigma(1775) \leftrightarrow \Lambda^0+\pi$ on the left
and $\Sigma(1750) \leftrightarrow \Sigma+\eta$ on the right; blue
and  red lines are for decay and backward fusion reaction,
respectively.}} \label{rates1775}
\end{figure*}

Particle fugacities, except $\Upsilon_{\pi}$, change rather rapidly.
Figure~\ref{Upsil} shows the computed $\Upsilon(t)$ as a function of
temperature $T(t)$. We present here the  scenario in which  all
reactions evolve in both directions, for the initial condition
$\gamma_s=\gamma_q$. The time, corresponding to the temperature
shown at the bottom, is shown at the top of figure~\ref{Upsil}, in
each frame. On the left we have hadronization at 140 MeV, in the
middle at 160 and to the right at 180 MeV. Each frame has the same
scale size for temperature unit, not time. For $\Upsilon_{\Sigma^*}$
we show two possible evolution examples, for $\Sigma_{1750}$
(dash-dot dark line) and $\Sigma(1775)$ (dashed line). These
resonances have significant influence on the $\Lambda(1520)$ yield.
The solid lines are for $\Upsilon_{\Lambda(1520)}$ (upper, red line)
and $\Upsilon_{\Sigma(1385)}$ (lower, light blue line). The dash-dot
and dashed light lines are for $\Upsilon_{\Sigma(1190)}$ and
$\Upsilon_{\Lambda^0}$, respectively. The upper dotted line is for
$\Upsilon_{N}$ and lower dotted line is for $\Upsilon_K$.

An important feature is that the $\Upsilon$s of massive hadron
(resonances) increase very fast when $T$ decreases.  This is so
since in absence of a rapid re-equilibration  reactions,
multiplicity of given resonance must  be conserved. Then, according
to Eq.~(\ref{relboltz}) $\Upsilon_i \propto 1/K_2(m_i/T)$, and thus
for large $m_i$ $\Upsilon_i \propto \exp(m_i/T)$. We would expect
$\Upsilon_i>\Upsilon_j$, when $m_i > m_j$, and $T$ decreases. This
behavior is just like we found for the case of large charm fugacity
~\cite{Kuznetsova:2006bh}. However, because of the decay and
regeneration reactions there are some deviations from this
expectation in figure~\ref{Upsil}.

For $T_0=180$ MeV in most cases $\Upsilon_3>\Upsilon_1\Upsilon_2$
$(t>0)$. Massive resonances decay to lower mass particles. The
result is defined by resonance mass, its decay width and decay
products. For example $\Upsilon_{\Sigma(1775)}$ is smaller  than
$\Upsilon_{\Sigma(1750)}$ and $\Upsilon_{\Lambda(1520)}$, because of
its large decay width. Therefore excitation of $\Sigma(1775)$ by
$\Lambda$ slightly dominates over $\Sigma(1775)$ decay to
$\Lambda(1520)$ even in this case, when for most resonances the
decay is dominant. For smaller initial hadronization temperatures
$\Upsilon_{\Lambda(1520)}$ becomes smaller than
$\Upsilon_{\Sigma(1775)}$, and  even smaller than
$\Upsilon_{\Sigma(1385)}$ in some range of temperatures. This
suppression occurs because of $\Sigma(1775)$, and others $\Sigma^*$
regeneration. Because of large $\Upsilon_{\pi}$,
$\Upsilon_{\Sigma(1775)} < \Upsilon_{\Lambda(1520)}\Upsilon_{\pi}$,
the $\Sigma(1775)$ production by $\Lambda(1520)$ is dominant in the
full range of $T$  considered here.

\subsection{Final $\Lambda(1520)$ and $\Sigma(1385)$ multiplicities}\label{chapter 3.2}
In this section  we consider the evolution of the multiplicity of \
resonances $\Lambda(1520)$, $\Sigma(1385)$, $\Sigma(1775)$ during
the kinetic phase. We use the Boltzmann yield limit,
Eq.\,(\ref{relboltz}). By the symbol $X(T)$  we refer to a
particular resonance, and $X_0$ is the initial multiplicity for that
resonance. The dynamic yield of this resonance may be expressed as
\begin{equation}
\frac{X(T)}{X_0} = \frac{\Upsilon_X(t)T(t)^3K_2(m_X/T(t))}
            {\Upsilon_{X\,0}T_0^3K_2(m_X/T_0)}
\end{equation}

Figure~\ref{Lam1520r} shows this yield as a function of T(t) for
$X=\Sigma(1385)$ (left) and $X=\Lambda(1520)$ (right). We consider
three   initial conditions, temperature $T_0 = 140, 160, 180$ MeV,
with corresponding $\gamma_q =1.6, 1.27, 1.0$, respectively. The
solid lines correspond for the model with dead channels and dashed
one are for case when all reactions are symmetric  in both
directions. The thin dotted vertical line at $T=120$ MeV marks the
kinetic freeze-out temperature, assumed before
in~\cite{Kuznetsova:2008zr}. The main result is that the resulting
relative yields for $\Lambda(1520)$ and $\Sigma(1385)$ behave
qualitatively different from  each other. In particular, as the
temperature decreases, for the case $T_0 =140$ MeV we observe a
strong yield suppression for $\Lambda(1520)$, and a strong
enhancement for $\Sigma(1385)$ (as compared to initial SHM yields).

To  better understand the mechanism of $\Lambda(1520)$ suppression,
we analyze in some detail the  case of $\Sigma(1775)$ and
$\Sigma(1750)$ decay and production rates $dW/dVdt$. We assume here
that these reactions can go in both directions. In figure
\ref{rates1775} we show the reactions rates for the principal
channels of decay and production as a functions of temperature T for
$\Sigma(1775)$ (left) and $\Sigma(1750)$ (right), for the case of
initial temperature  $T_0=140$ MeV which provides the largest
$\Lambda(1520)$ suppression. Solid lines are for the reaction
$\Sigma \leftrightarrow \Lambda(1520) + \pi$, dash-dot lines are for
reaction $\Sigma \leftrightarrow N+K$, dashed lines are for reaction
$\Sigma \leftrightarrow \Lambda^0+\pi$. Two set of lines are
presented for the  decay (on-line blue) and backward fusion reaction
(on-line red), respectively.

As temperature decreases, all rates $dW/dtdV$ are increasing
rapidly. This is mainly because fugacities $\Upsilon$  increase
nearly exponentially when number of particles is conserved, see
figure \ref{Upsil}. We see that at the beginning of the kinetic
phase all reactions go in the direction of $\Sigma(1775)$
production, since $\Sigma(1775)$ production rate is larger than its
decay rate for all channels. Then at  first $\Sigma(1775)
\leftrightarrow \Lambda^0+\pi$   decay rate    becomes dominant over
$\Sigma(1775)$ production rate in this channel, followed by the same
for $\Sigma(1775)\leftrightarrow N+K$ channel.

For the reaction  $\Sigma(1775) \leftrightarrow \Lambda(1520) + \pi$
backward reaction is always dominant. As  result, during the
kinetic phase always more $\Lambda(1520)$ resonances are excited
into $\Sigma(1775)$ than they are produced by $\Sigma(1775)$ decay.
The reason for this is the decay of $\Sigma(1775)$ to the other channels,
as long as   $\Upsilon_{\Sigma(1775)}<\Upsilon_{\Lambda(1520)}\Upsilon_{\pi}$.
The lighter is the total mass of decay products,  the earlier the
decay reaction becomes dominant. This is due to the fact that
the fugacity of $\Upsilon$ for heavier particles increases faster
with expansion. Therefore, the decay rate becomes dominant earlier,
when the difference between initial and final mass is larger.
The  net result is $\Lambda(1520)$ suppression by $\Sigma(1775)$ excitation.

\begin{figure}
\centering
\includegraphics[width=8 cm, height=10 cm]{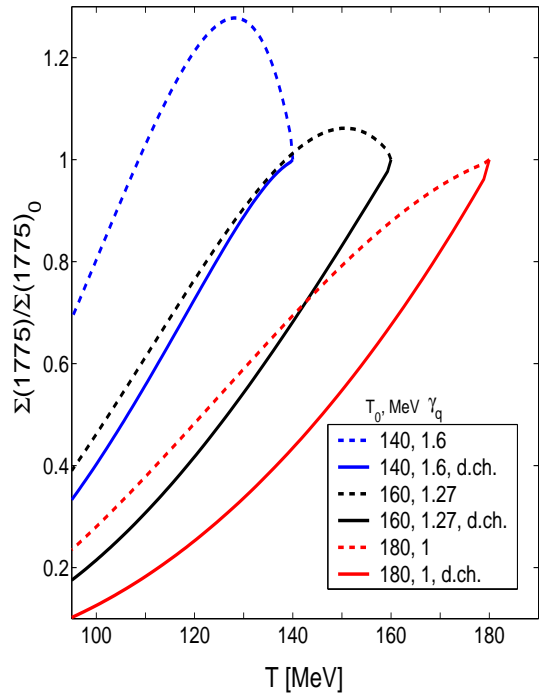}
\caption{\small{The ratio
$\Sigma(1775)/\Sigma(1775)_0$
as a functions of temperature $T(t)$ for
different initial hadronization temperatures $T_0=140$, $160$
and $180$ MeV (blue/bottom, black/middle and red/top lines), respectively. Solid lines
are for calculations with dead channels, dashed lines are for
calculations without dead channels.}} \label{Sigm1775r}
\end{figure}

In figure~\ref{Sigm1775r} we show the yield of $\Sigma(1775)$
normalized by its initial yield at hadronization:
$\Sigma(1775)/\Sigma(1775)_0$ as a function of $T(t)$. Like in the
other figures above, solid lines are for the dead channels and
dashed lines are for case when reactions go in both directions,
solid (blue) lines are for $T_0=140$ MeV, solid (black) lines for
$T_0=160$ MeV, and solid (red) lines are for $T_0=180$ MeV. Each of
the lines can be identified by their initial $T$-value. We see that
when all reactions go in both direction the ratio
$\Sigma(1775)/\Sigma(1775)_0$ increases at first similar to
$\Sigma(1385)/\Sigma(1385)_0$ and $\Delta(1230)/\Delta(1230)_0$
ratios~\cite{Kuznetsova:2008zr}.

Compared to these ratios, $\Sigma(1775)/\Sigma(1775)_0$ ratio
reaches its maximum value earlier, and after the maximum, the yield
of $\Sigma(1775)$ decreases faster. The reason for this behavior is
that the mass of $\Sigma(1775)$ is larger. The phase space occupancy
$\Upsilon_{\Sigma(1775)}$, and therefore its decay rates,  increase
faster than the fugacity and decay rates for $\Sigma(1385)$ and
$\Delta(1230)$. Therefore decays $\Sigma(1775)$ to some channels and
its total decay rate become dominant earlier (see figure
\ref{rates1775}). Although the total decay width of $\Sigma(1775)$
is approximately the same as for $\Delta(1230)$, the maximum value
of this ratio is smaller.

Said differently, the maximum yield of $\Sigma(1775)$ does not have
time to reach the value as high as that for $\Delta(1230)$. We thus
learn that the time evolution of the yield of resonances with large
decay width depends not only on their decay width, but also on mass
difference between initial and final states. Similar time evolution
occurs for the other $\Sigma^*$, which quantitatively depends on
their mass,
 decay products masses and decay width.

For most $\Sigma^*$s, the decay products in the channel
$\Lambda(1520)+\pi$ are heavier than the decay products in others
channels, which are thus favored by phase space.  For most
resonances in our range of temperature, the decay into
$\Lambda(1520)+\pi$ remains weak. The exception is $\Sigma(1750)$
which decays also to $\Sigma + \eta$, see figure~\ref{rates1775}.
($m_{\Sigma}+m_{\eta} > m_{\Lambda(1520)}+m_{\pi}$). $\Sigma(1750)$
begins to decay dominantly to $\Lambda^0(1520)$ at relatively low
temperature $T=116$ MeV, and continues to be produced by
$\Sigma+\eta$ fusion.

As a result,  allowing  all reactions to go in both directions, the
ratio $\Lambda(1520)/\Lambda(1520)_0$ has a minimum. This is
specifically due to  $\Sigma(1750)$) decay back to $\Lambda(1520)$
at small temperatures as described above. However,  when we  satisfy
Eq.(\ref{dchcon})  for dead channels the only decay occurs in the
beginning of kinetic the dead-channel model  phase.
 In that case the $\Upsilon_{\Sigma^*}$s are smaller,
and the rate of reaction $\Lambda(1520)+\pi\rightarrow \Sigma^*$
exceeds the rate for backward reaction by  larger amount, compared
to the scenario without dead channels. This amplifies the effect of
$\Lambda(1520)$ suppression. In this case, $\Sigma^*$  decay to
lighter hadrons right after they are produced by $\Lambda(1520)$. We
can see that for $T_0 = 140$ MeV and $T_0 = 160$ MeV $\Lambda(1520)$
yield is always decreasing in the here considered temperature range.

For $\Sigma(1385)$ multiplicity we find a result quite different from
$\Lambda(1520)$ behavior discussed here, but  similar to what we obtained in
\cite{Kuznetsova:2008zr} by a very different method in a smaller basis set of states.
In particular, the $\Sigma(1385)$
yield is enhanced, but the maximum value of
$\Sigma(1385)/\Sigma(1385)_0$ we find is  a few percent  higher, since
 we took into account the Bose enhancement of interaction rates,
reaction (\ref{S1385S}), and $\Sigma^*$ production. $\Sigma(1385)$ contribution to
$\Sigma^*$ production is small, compared to the influence of the first two effects.
The time (i.e. temperature) evolution of $\Sigma(1385)$ practically does not depend on
the presence of dead channels, and the   maximum enhancement of $\Sigma(1385)$
is even less sensitive. This in fact
indirectly confirms that  $\Sigma^*$ has a small influence on $\Sigma(1385)$ multiplicity.
Thus we confirm that: \\
a)   for $T_0=180$ MeV $\Sigma(1385)$   evolves with the system following
the ambient temperature;\\
b)    for $T_0=160$ MeV $\Sigma(1385)$  shows some increase in yield;\\
c)   for $T_0=140$ MeV there is a strong yield increase of
$\Sigma(1385)$.

While there is little sensitivity in the  yield of $\Sigma(1385)$ to issue of particle momentum distribution
 (little difference between the two models considered, dashed and solid lines), the $\Sigma(1385)$   yield is highly
sensitive to initial  hadronization condition.  While for
$\Sigma(1385)$ the yield increases with decreased hadronization
temperature, for $\Lambda(1520)$ the opposite is true,  and in
particular the smallest final $\Lambda(1520)$ yield  corresponds to
the smallest hadronization temperature  for both models.

\begin{figure*}
\centering
\includegraphics[width=8 cm, height=10 cm]{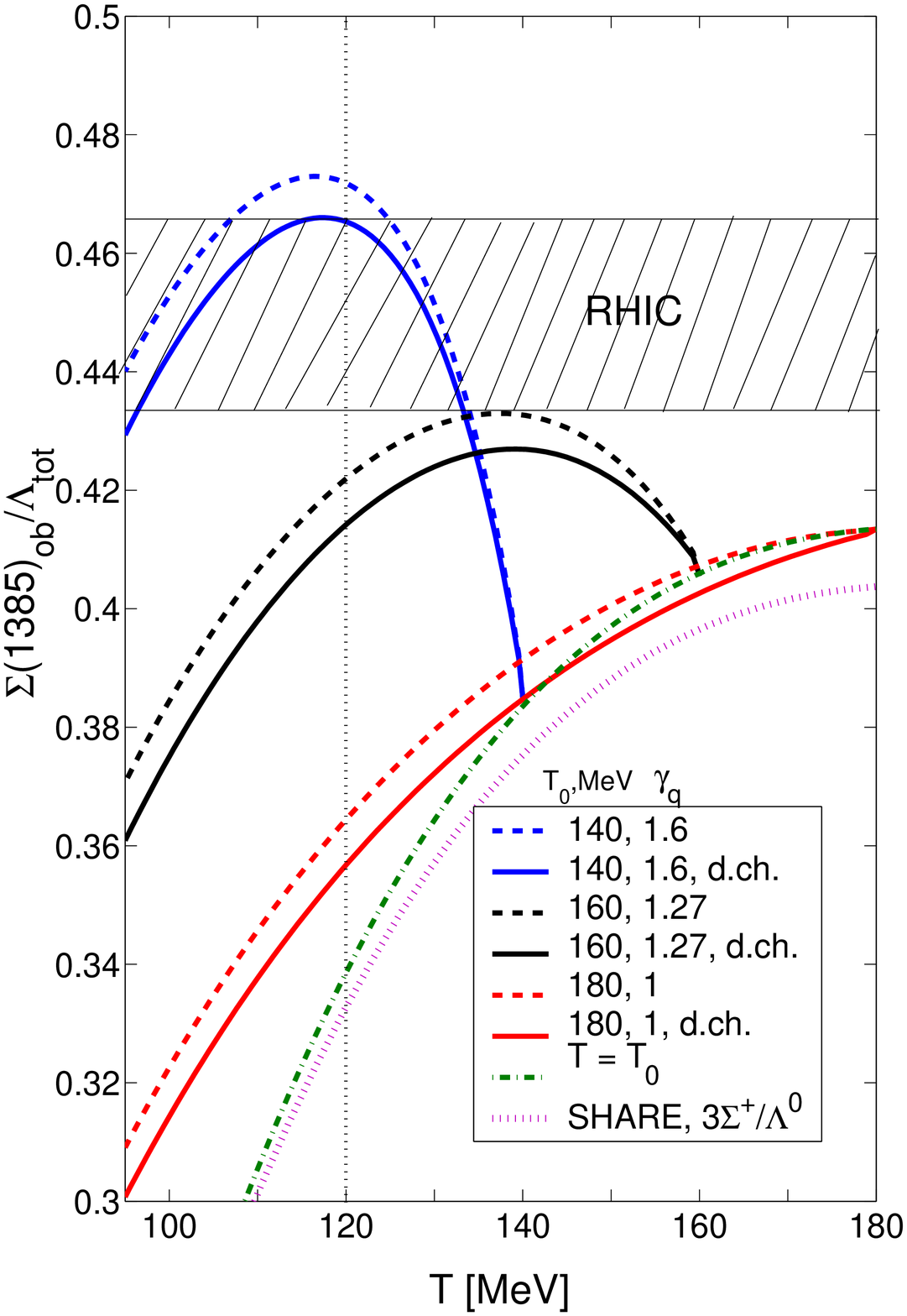}
\includegraphics[width=8 cm, height=10 cm]{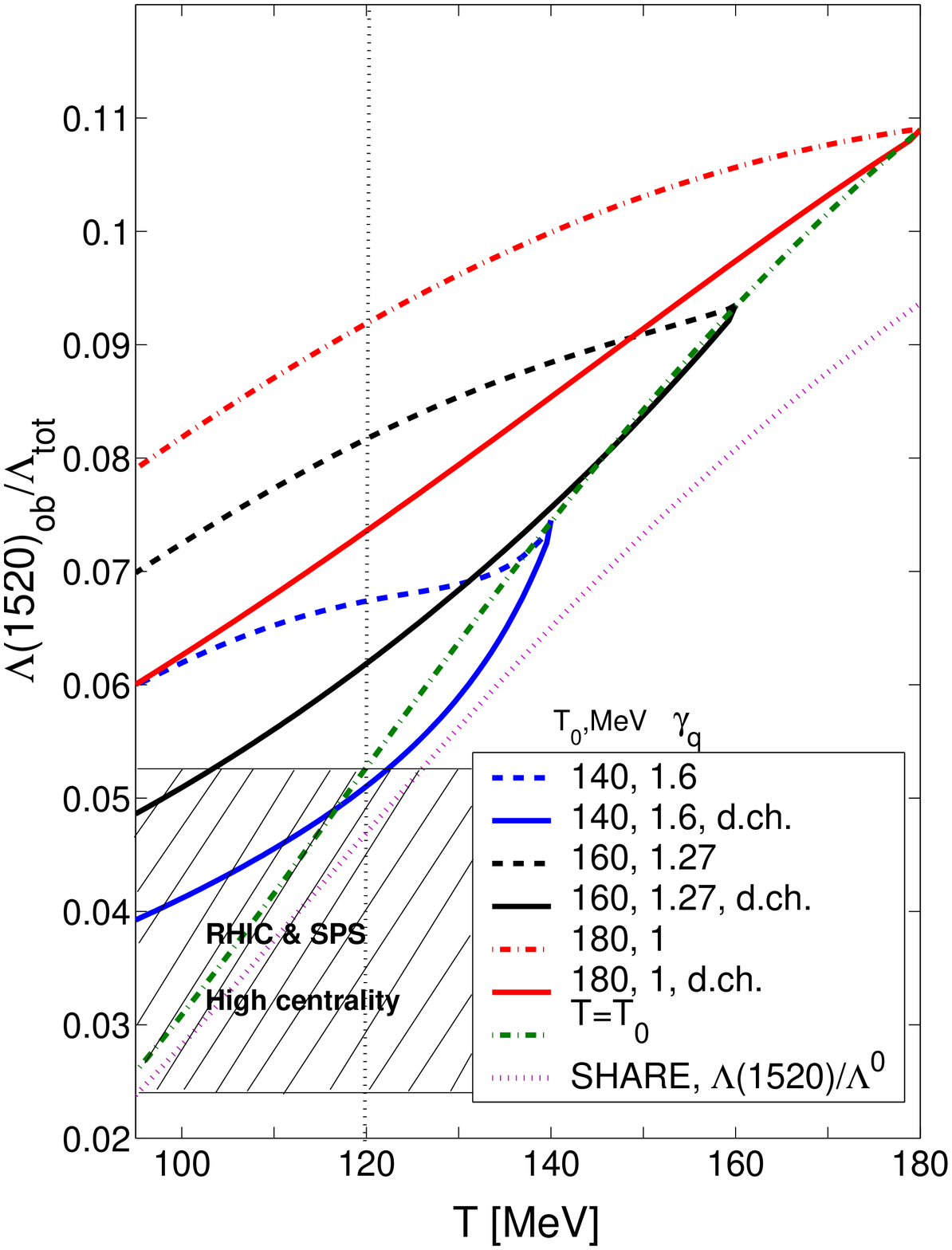}
\caption{\small{The ratios $\Sigma(1385)/\Lambda_{\rm tot}$ (on
left) and $\Lambda(1520)/\Lambda_{\rm tot}$ (on right) as a function
of temperature $T$ of final kinetic freeze-out, for different
initial hadronization temperatures $T_0=140$, $160$ and $180$ MeV
(blue, black and red lines, respectively). Dashed lines are for
calculations without dead channels, solid lines are for calculations
with dead channels.  The dotted purple line gives the expected SHM
chemical equilibrium  result. The dash-dot line is relative yield
result result for SHM with $T_0=T$.}} \label{lamltot}
\end{figure*}

\subsection{Experimentally measurable resonance  ratios}\label{chapter 3.3}

The initial hadronization yields, which we used as a reference in
figure~\ref{Lam1520r}  in order to understand the physical behavior,
are not measurable. What is commonly used as a reference for the
yields of single strange hyperon resonances is the overall yield of
the stable $\Lambda^0(1115)$, without the weak decay feed from
$\Xi$. Aside of the initially produced particles, the experimental
yield of $\Lambda^0(1115)$  also includes resonances decaying during
the free expansion after kinetic freeze-out, in particular (nearly)
all decays of   $\Sigma(1385)$, and the experimentally inseparable
yield of $\Sigma^0(1193) \to \gamma+\Lambda^0$ decay and the decay
of any further hyperon resonances $Y^*$.

Thus we normalize our final result with the experimentally
observable final $\Lambda_{\rm tot}^0$ hyperon yield:
\begin{equation} \label{Ltot}
\Lambda_{\rm tot}= \Sigma^0(1193)  + 0.91 \Sigma(1385)+ \Lambda + Y^*.
\end{equation}
The factor $0.91$ shows that $91\%$ of end-state $\Sigma (1385)$
decays to $\Lambda$. We also included in  $\Lambda_{\rm tot}$
calculations decays of $\Xi^* \rightarrow \Lambda+K$, which makes
the result slightly dependent on $\gamma_s/\gamma_q$ ratio. We use
$\gamma_s/\gamma_q =1$, since this ratio value is expected at top
RHIC energy~\cite{Kuznetsova:2006bh}.

As noted,   $\Lambda(1520)$ and $\Sigma(1385)$
experimentally observable yields also include any decays which occur
in the free-streaming post-kinetic period. Thus we have:
\begin{eqnarray}
\Sigma(1385)_{\rm ob}=\Sigma(1385)+Y^*_{\Sigma(1385)},\\
\Lambda(1520)_{\rm ob}=\Lambda(1520)+Y^*_{\Lambda(1520)},
\end{eqnarray}
where $Y^*_{\Sigma(1385)}$ and $Y^*_{\Lambda(1520)}$ are hyperon
multiplicities at kinetic freeze-out temperature, and which decay to
$\Sigma(1385)$ and $\Lambda(1520)$, respectively. The multiplicities
$\Sigma(1385)$ and $\Lambda(1520)$ are taken at the moment of
kinetic freeze - out.

In  figure~\ref{lamltot} we present the fractional yields
${\Sigma(1385)}/\Lambda_{\rm tot}$ (left), and
${\Lambda(1520)}/\Lambda_{\rm tot}$ (right) as a function
of temperature of final kinetic freeze-out $T$.
The results  for the hadronization
temperatures $T_0=140$ (blue lines), $T_0=160$ (black
lines) and $T_0=180$ MeV (red lines) are shown. Solid lines
are for the case with dead channels and dashed lines are for the
case when all reactions are going in both directions.

In  figure~\ref{lamltot}  the green dash-dotted line is the result
when the kinetic freeze-out temperature $T$ coincides with the
hadronization temperature $T_0$. There is no kinetic phase in this
case, only   resonances decay after hadronization. This result is
similar to SHARE result (purple, dotted line). The  small difference
is mainly due to us taking into account the decays
\begin{equation}
\Sigma(1670, 1750) \rightarrow \Lambda(1520) + \pi,
\end{equation}
which are expected/predicted in~\cite{Cameron:1977jr}.
Similarly, for $\Sigma(1385)$ our results for $T_0=T$
are different from SHARE results because we include the decay:
\begin{equation}
\Sigma(1670) \rightarrow \Sigma(1385) + \pi,
\end{equation}
expected/predicted in~\cite{Prevost:1974hf}. These additional resonances
are part of current particle data set~\cite{Amsler:2008zz}.

For all initial
hadronization temperatures, as the freeze-out temperature decreases, the suppression for
$\Lambda(1520)_{\rm ob}/\Lambda_{\rm tot}$ ratio is larger than for
$\Lambda(1520)/\Lambda(1520)_0$ (at the same temperature $T$ of final kinetic freeze-out).
 This is particularly evident for dead channels and hadronization temperatures $T_0=160,\, 180$ MeV
(see figure~{\ref{Lam1520r}}).
The effect is due to $\Sigma(1775)$ suppression, as shown in
figure~\ref{Sigm1775r} (and similar for other $\Sigma^*$).
For $T_0=140$ MeV the additional suppression of $\Lambda(1520)$,
described above, is relatively small.

For $T_0=140$ MeV in the case without dead channels at final kinetic
freeze-out $T>120\,{\rm MeV}$, the final observed $\Lambda(1520)$
suppression is even smaller, compared to its suppression in the
kinetic phase at the same temperature  (see
figure~{\ref{Lam1520r}}). The reason is that yield of $\Sigma(1775)$
(and of the other $\Sigma^*$s) is much enhanced for this range of
temperatures see figure~\ref{Sigm1775r}. This additional
$\Sigma(1775)$ decays back to $\Lambda(1520)$. That results in a
smaller suppression at these temperatures.

The  above suppression effect   increases in magnitude  for higher
hadronization temperatures, since the suppression of $\Sigma(1775)$
and the sensitivity of $\Lambda(1520)_{\rm ob}$ multiplicity to
$\Sigma^*$ decays increase with temperature. However, when we
consider dead channels (see figure~\ref{lamltot}), the former effect
of $\Lambda(1520)$ suppression during evolution of  kinetic phase
increases for decreasing hadronization temperatures. Thus in the
combined effect, the  observable  relative suppression of
$\Lambda(1520)_{\rm ob}/\Lambda_{\rm tot}$, is approximately of the
same magnitude  for all hadronization temperatures $T_0$. However,
the initial hadronization yield of  $\Lambda(1520)$ is sensitive to
temperature, and decreases rapidly with $T$. Therefore only for $T_0
= 140 $ MeV, a kinetic freeze-out temperatures $\approx 95 -105$
MeV, and allowing for dead channels, the  ratio $\Lambda_{\rm
ob}(1520)/\Lambda_{\rm tot}$ reaches the experimental domain
$\Lambda_{\rm ob}(1520)/\Lambda_{\rm tot}<0.042\pm
0.01$~\cite{Adams:2006yu,Markert:2002xi} shown in
figure~\ref{lamltot} by dashed lines.

For the same initial conditions, that is
for $T_0=140$ MeV, we find the ratio $\Sigma(1385)/\Lambda_{\rm tot} \approx$
 0.45 at $T \approx 100$ MeV  (and for the entire range 95 -- 135 MeV,
in good agreement with experimental
data~\cite{Adams:2006yu,Salur:2006jq}). In~\cite{Kuznetsova:2008zr}
this value of $\Sigma(1385)/\Lambda_{\rm tot}$  is found at $T=120$ MeV,
which was in the reference the presumed lowest possible temperature of the final kinetic freeze-out.
Here we find that at $T=120$ MeV the ratio $\Sigma(1385)/\Lambda_{\rm tot}$ can be even higher
(about 0.47), which is due to the Bose enhancement of
in-medium $\Sigma(1385)$ production rate (see discussion following
figure~\ref{Lam1520r}).

\section{Conclusions}\label{chapter 4}
The resonant hadron states, considering their very  large decay and
reaction  rates, can interact beyond the chemical and thermal
freeze-out of stable particles. Thus the observed yield of
resonances is fixed by the physical conditions prevailing at a later
breakup of the fireball matter rather than the production of
non-resonantly interacting hadrons. Moreover,  resonances, observed
in  terms of the invariant mass signature,  are only visible when
emerging from a more dilute hadron system  given the ample potential
for rescattering of decay products. The combination of experimental
invariant mass method with a large resonant scattering makes the
here presented  population study of resonance kinetic  freeze-out
necessary. The evolution effects we find are greatly amplified at
low hadronization temperatures where greatest degree of initial
chemical equilibrium is present.

Our study quantifies the expectation that in a dense hadron  medium
narrow resonances are ``quenched''\cite{Rafelski:2001hp} that is,
effectively mixed with other states, and thus their observed
population is reduced. Since we follow here the particle density,
the effect we study is due to incoherent population  mixing of
$\Lambda(1520)$, in particular with $\Sigma^*$. This effect is
possible for particle densities out of chemical non-equilibrium.
However, this mixing can occur also at the amplitude (quantum
coherent) level. As the result the yield suppression effect  could
further increase, in some situations further improving the agreement
with experiment.

Our results show that the observable ratio $\Lambda(1520)_{\rm
ob}/\Lambda_{\rm tot}$ can be suppressed by two effects. First
$\Lambda(1520)$ yield is suppressed   due to excitation of heavy
$\Sigma^*$s in the resonance scattering process. Moreover, the final
$\Lambda(1520)_{\rm ob}$ yield is suppressed, because $\Sigma^*$s,
which decay to $\Lambda(1520)$, are suppressed at the end of the
kinetic phase evolution by their (asymmetric) decays to lower mass
hadrons, especially when dead channels are present (see
figure~\ref{Sigm1775r}). As a result, fewer  of these hadrons can
decay to $\Lambda(1520)_{\rm ob}$ during the following free
expansion. A contrary mechanism operates for the resonances such as
$\Sigma(1385), \Delta(1230)$. These resonances can be so strongly
enhanced, that in essence most final states strange and non-strange
baryons come from a resonance decay.

We note that despite a   scenario dependent resonance formation or suppression,
the stable particle yields used in study of chemical freeze-out remain unchanged, since
all resonances ultimately decay into the lowest ``stable'' hadron.   Therefore after a
description e.g. within a statistical hadronization model  of the yields
of stable hadrons, the understanding of resonance yields is a second, and  separate task
which helps to establish the consistency of our physical understanding of the hadron
production process.

We conclude noting the key result of this study, that   we can now understand the opposite behavior
of $\Lambda(1520)$ (suppression in high centrality reactions) and $\Sigma^(1385) $ (enhancement,
 and similarly $\Delta(1230)$) by considering their rescattering in matter. In order to explain both,
the behavior of the $\Lambda(1520)_{\rm ob}/\Lambda_{\rm tot}$
and $\Sigma(1385)/\Lambda_{\rm tot}$ ratios, one has to consider
$T=95-100$ MeV as the  favorite temperature of final kinetic
freeze-out  of hadron resonances, with   $T_0=140$ MeV being the
favored chemical freeze-out (hadronization, QGP break-up) temperature.
When there is little  matter  available to scatter, e.g. in peripheral  collisions,
  the average value of
$\Lambda(1520)_{\rm ob}/\Lambda_{\rm tot}$ ratio is higher,
approaching  the   expected chemical freeze-out  hadronization yield
for $T_0=140$ MeV. All these findings are in good agreement with
available experimental data.

\subsubsection*{Acknowledgments}
This research was supported
by a grant from: the U.S. Department of Energy  DE-FG02-04ER4131;
and  by the DFG-LMU   Excellent program.



\vspace*{-0.3cm}

\end{document}